\documentclass[8.5pt,twocolumn]{article}
\usepackage[super,sort&compress,comma]{natbib} 
\usepackage{mhchem}
\usepackage{times,mathptmx}
\usepackage{sectsty}
\usepackage{balance} 

\usepackage{graphicx} 
\usepackage{lastpage}
\usepackage[format=plain,justification=raggedright,singlelinecheck=false,font=small,labelfont=bf,labelsep=space]{caption} 
\usepackage{fancyhdr}
\usepackage{color}

\usepackage{amsmath}
\usepackage{amssymb}
\usepackage{amsthm}
\usepackage{array}
\usepackage[pdfnewwindow]{hyperref}
\usepackage{bm}
\pdfoutput=1
\def\kt{k_{\rm B}T}

\begin{document}

\thispagestyle{plain}
\fancypagestyle{plain}{
\renewcommand{\headrulewidth}{1pt}}
\renewcommand{\thefootnote}{\fnsymbol{footnote}}
\renewcommand\footnoterule{\vspace*{1pt}%
\hrule width 3.4in height 0.4pt \vspace*{5pt}} 
\setcounter{secnumdepth}{5}

\makeatletter 
\def\subsubsection{\@startsection{subsubsection}{3}{10pt}{-1.25ex plus -1ex minus -.1ex}{0ex plus 0ex}{\normalsize\bf}} 
\def\paragraph{\@startsection{paragraph}{4}{10pt}{-1.25ex plus -1ex minus -.1ex}{0ex plus 0ex}{\normalsize\textit}} 
\renewcommand\@biblabel[1]{#1}            
\renewcommand\@makefntext[1]%
{\noindent\makebox[0pt][r]{\@thefnmark\,}#1}
\makeatother 
\renewcommand{\figurename}{\small{Fig.}~}
\sectionfont{\large}
\subsectionfont{\normalsize} 

\fancyhead{}
\renewcommand{\headrulewidth}{1pt} 
\renewcommand{\footrulewidth}{1pt}
\setlength{\arrayrulewidth}{1pt}
\setlength{\columnsep}{6.5mm}
\setlength\bibsep{1pt}

\twocolumn[
  \begin{@twocolumnfalse}
\noindent\LARGE{\textbf{Do hierarchical structures assemble best via hierarchical pathways?}}
\vspace{0.6cm}

\noindent\large{\textbf{Thomas K. Haxton and Stephen Whitelam}}

\noindent{\textit{Molecular Foundry, Lawrence Berkeley National Laboratory, Berkeley, CA 94720, USA}}
\vspace{0.5cm}



\noindent \normalsize{Hierarchically structured natural materials possess functionalities unattainable to the same components organized or mixed in simpler ways. For instance, the bones and teeth of mammals are far stronger and more durable than the mineral phases from which they are derived because their constituents are organized hierarchically from the molecular scale to the macroscale. Making similarly functional synthetic hierarchical materials will require an understanding of how to promote the self-assembly of structure on multiple lengthscales, without falling foul of numerous possible kinetic traps. Here we use computer simulation to study the self-assembly of a simple hierarchical structure, a square crystal lattice whose repeat unit is a tetramer. Although the target material is organized hierarchically, it self-assembles most reliably when its dynamic assembly pathway consists of the sequential addition of monomers to a single structure. Hierarchical dynamic pathways via dimer and tetramer intermediates are also viable modes of assembly, but result in general in lower yield:  these intermediates have a stronger tendency than monomers to associate in ways not compatible with the target structure. 
In addition, assembly via tetramers results in a kinetic trap whereby material is sequestered in trimers that cannot combine to form the target crystal.  
We use analytic theory to relate dynamical pathways to the presence of equilibrium phases close in free energy to the target structure, and to identify the thermodynamic principles underpinning optimum self-assembly in this model: 1) make the free energy gap between the target phase and the most stable fluid phase of order $k_{\rm B} T$, and 2) ensure that no other dense phases (liquids or close-packed solids of monomers or oligomers)
or fluids of incomplete building blocks 
fall within this gap.}
\vspace{0.5cm}

 \end{@twocolumnfalse}
 ]




\section{Introduction}

Since assembly pathway strongly determines the ability of structures to self-assemble efficiently, it is vital to understand what rules link building block design, assembly pathway, and assembly yield.  While some structures assemble via two-step mechanisms involving intermediate structures distinct from both the parent and target structures~\cite{vekilovtwo}, an important dichotomy separates even those pathways whose intermediates remain commensurate with the target structure: assembly may occur either \textit{directly} via sequential accumulation of monomers, as in classical theories of nucleation and growth~\cite{becker1935kinetische, volmer1926keimbildung}, or \textit{hierarchically} via formation of intermediate-scale structures that act as the building blocks for subsequent stages of assembly.  In general, hierarchical pathways dominate whenever clusters meet via diffusion faster than they exchange monomers~\cite{whitelam2009role}, so in principle any system's assembly pathway can be adjusted by tuning binding energies~\cite{PhysRevLett.102.118106}.  For instance, the pathway for phase separation of binary fluids changes with increasing supercooling from direct assembly to hierarchical coalescence of clusters~\cite{bray1994theory, cates2000soft}.  While binary fluids demix more efficiently via their direct pathway, it seems reasonable to expect that materials with hierarchical \textit{structure} might assemble more efficiently via hierarchical pathways.  Nature abounds with materials such as collagen~\cite{Fallas2010}, enamel~\cite{Fang2011}, virus capsids~\cite{Zlotnick2011}, cytoskeletal structures~\cite{Herrmann2004}, and biominerals~\cite{mann2001biomineralization, dove2003biomineralization}, whose functionality depends on their having structure on multiple length scales~\cite{fratzl2007nature}.  Our pursuit of similarly functional hierarchical materials~\cite{li2012direction, liao2012real, Murnen2010, Miszta2011, Tao2012, Rehm2012} would be aided by knowing which dynamic modes of assembly to promote, and which to suppress.

Here we ask the question: do hierarchical structures assemble best via hierarchical pathways?  We can begin to answer this question through the intensive study of computer models inspired by real systems.  A simple example of a natural hierarchical material is provided by the SbpA surface-layer (S-layer) protein, which forms a membrane on the outsides of the bacterium \textit{Lysinibacillus sphaericus} and on surfaces {\em in vitro}~\cite{pum1991role, messner1992crystalline, sleytr1997baa, chung2010self}. This membrane, which is robust and controllably porous, is a square crystal lattice whose repeat unit is a tetramer, and so is a member of perhaps the simplest class of hierarchical materials, a one-component structure possessing order on two length scales (that of the monomer and that of the tetramer). Here we use computer simulation to study the self-assembly of a model of this system, broadly varying model parameters to alternately favor hierarchical or direct assembly pathways, in order to determine which dynamical pathways result in most reliable assembly of the target material. We also use analytic mean-field theory to relate these dynamical pathways to the model's underlying thermodynamics, thereby establishing a set of simple principles that describe where in phase space optimal self-assembly occurs. 

\begin{figure}[t]
\begin{center}
\includegraphics[width=\linewidth]{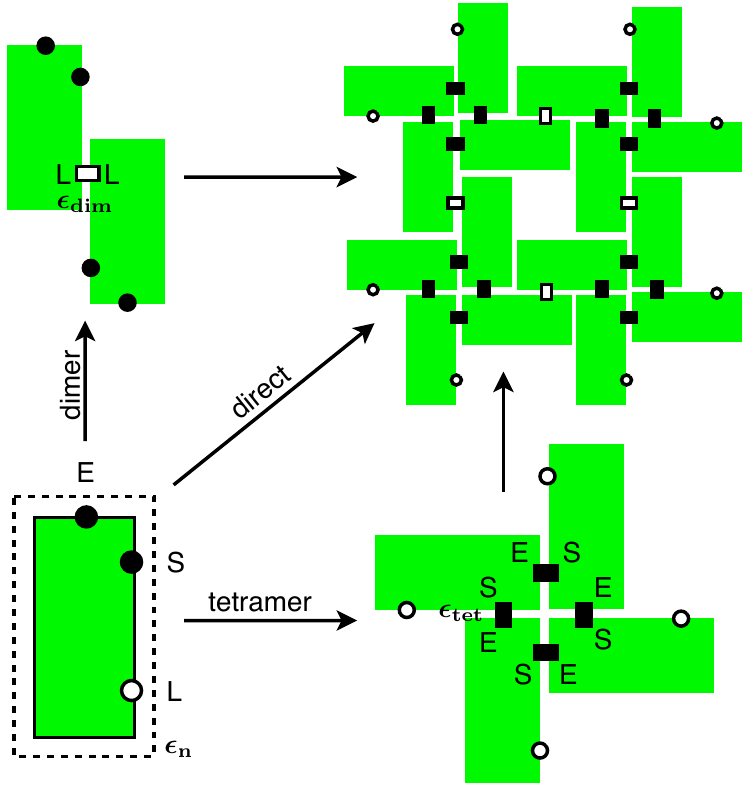}
\end{center}
\caption{{\em Model geometry.} Two chemically specific attractive interactions, a tetramer-stabilizing interaction $\epsilon_{\rm tet}$ between patches of type `E' and `S', and a dimer-stabilizing interaction $\epsilon_{\rm dim}$ between patches of type `L,' make possible the self-assembly of monomers (lower left) into an open square lattice of tetramers (upper right).  Specific interaction strengths may be tuned to favor different self-assembly pathways: non-hierarchical assembly involving the association of monomers (diagonal), assembly via the formation of tetramers (lower right), and assembly via the formation of dimers (upper left).  In addition, a nonspecific attractive interaction $\epsilon_{\rm n}$ (range depicted by dashed rectangle, lower right) allows the unstructured aggregation of monomers, dimers, and tetramers, and makes possible extended liquid and solid close-packed phases.}
\label{model}
\end{figure}

The constituent monomers of our model associate in three distinct ways (Fig.~\ref{model}): via chemically and directionally specific interactions that promote the formation of tetramers; via specific interactions that bind tetramers together (and which on their own promote the formation of dimers); and via nonspecific interactions that promote the formation of extended liquid and close-packed solid phases. Varying the strengths of the two specific interactions leads to three types of pathways for the assembly of the target structure: when the tetramer-forming interaction is strong, assembly occurs hierarchically via the formation of the tetramer repeat unit; when the dimer-forming interaction is strong, assembly occurs hierarchically via the formation of dimers; and when the two interaction strengths are balanced (in a roughly 1:2 ratio), assembly occurs non-hierarchically via the association of monomers.

In a previous paper~\cite{Haxton2012slayer} we focused on the non-hierarchical regime and determined how to optimize assembly by adjusting the balance of specific and nonspecific interactions. We found that assembly is best when 1) the free energy gap between the target phase and the most stable fluid phase is of order $k_{\rm B} T$; and 2) monomers associate in a weak nonspecific fashion, without promoting liquid-vapor phase separation. Although the target structure is an extended crystalline one, its self-assembly is in important ways more like the self-assembly of model virus capsids~\cite{sweeney2008exploring,hagan2006dynamic,rapaport2008role,nguyen2009invariant} -- whose components require a balance of interaction strength and specificity in order to associate reliably -- than the crystallization of spherical colloids, because the latter form of assembly {\em is} facilitated by liquid-liquid criticality~\cite{ten1997enhancement} or demixing~\cite{xu2012homogeneous}.

In this paper we vary all three interaction energies in order to determine what additional design principles apply when hierarchical assembly pathways are possible. The two design principles 
established in our previous work continue to hold. We also find that despite the hierarchical structure of the target material, optimal assembly occurs non-hierarchically, via sequential addition of individual building blocks to a larger structure. Assembly via dimer or tetramer intermediates leads in general to assembly of reduced quality, even though these intermediates are commensurate with the target structure. This reduction in quality stems from two types of kinetic trapping. The first involves the nonspecific aggregation of dimers or tetramers, which, because of the larger energy scales involved, occurs more readily than the nonspecific aggregation of monomers. The second, which occurs along tetramer-dominated pathways, involves the rapid formation of trimers (and the simultaneous depletion of the monomer pool), which cannot be combined to produce the target structure. This `incomplete building block' kinetic trap is also seen in computer simulations of virus capsid assembly~\cite{hagan2006dynamic,Zlotnick1999,Zlotnick2003,Hagan2010,Hagan2011}. We use analytic theory to identify the thermodynamics underpinning these traps, enabling us to refine design rule 2) for this model system: assembly is poor if liquid, close-packed solid, or incomplete-building-block phases lie lower in free energy than the parent fluid and within about $k_{\rm B} T$ of the stable square lattice.

\section{Model and methods}

\begin{figure*}
\begin{center}
\includegraphics[width=\linewidth]{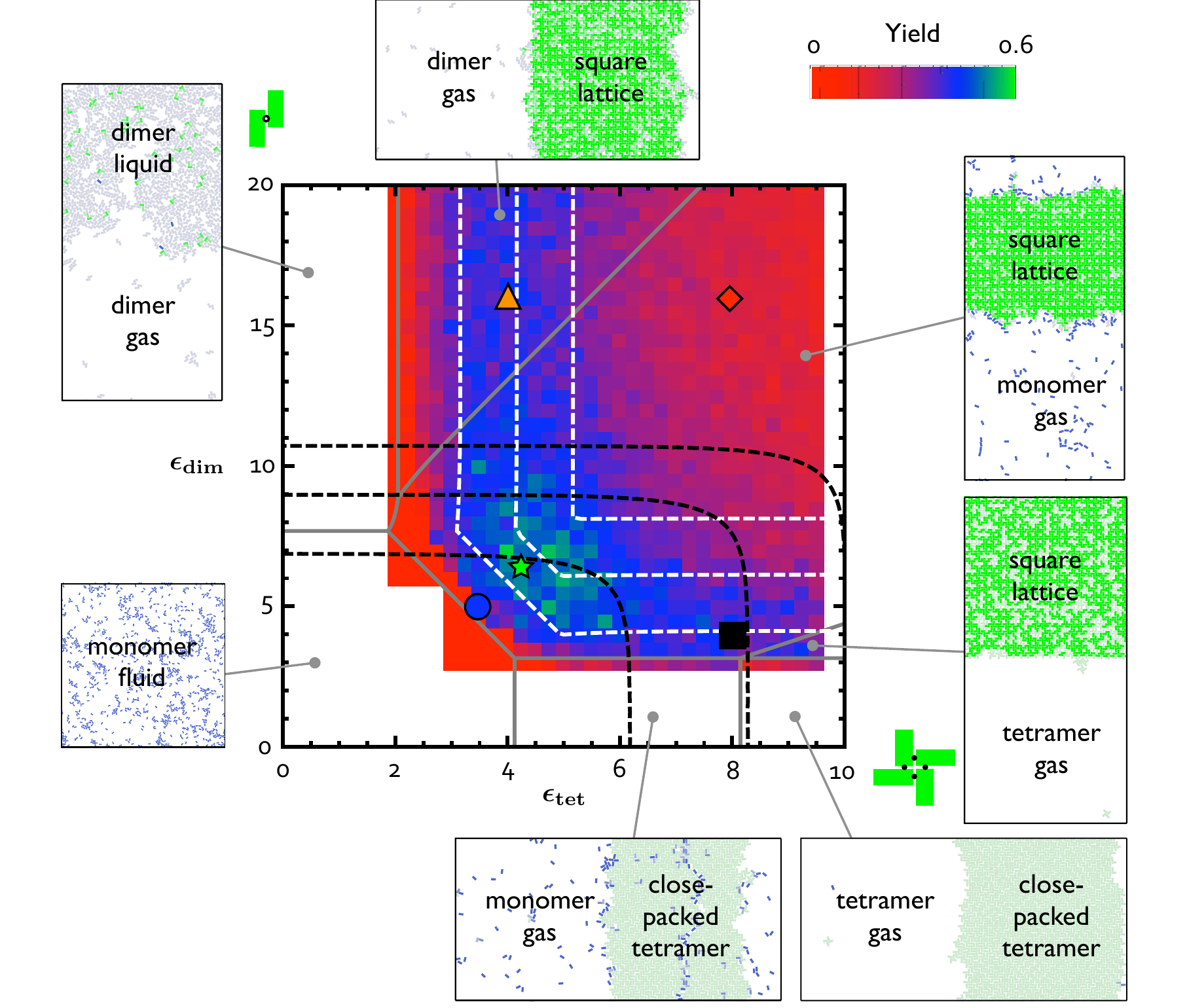}
\end{center}
\caption{\textit{Good assembly is localized in phase space.}  Phase diagram and dynamic yield after $10^7$ MC cycles as a function of specific interaction strengths $\epsilon_{\rm tet}$ and $\epsilon_{\rm dim}$ for a concentration $\phi=0.1$ and one choice of nonspecific interaction strength, $\epsilon_{\rm n}=2$.  Solid grey curves denote mean-field phase boundaries for the labeled simulated coexistence combinations.  The dashed black (white) curves denote lines of constant $B_2^\star$ (driving force $\mathcal{F}$).  Peak yield is localized and does not fall within a slot of constant $B_2^\star$.  A line of constant driving force $\mathcal{F}\approx \kt$ is a better predictor of (and a necessary condition for) good assembly. For this set of parameters, yield is best in the non-hierarchical sector (bottom left of the assembly region). The symbols connect with their counterparts in Figs.~\ref{snapshots} and \ref{timecourses}.}
\label{example}
\end{figure*}

We used the model studied in Refs.~\cite{Whitelam2010, Haxton2012slayer}, illustrated in Fig.~\ref{model}. Hard rectangular monomers of width $a$ and length $la$ ($l=11/5$) live on a smooth, two-dimensional substrate.  Monomers are designed to form a square lattice whose repeat unit is a tetramer. The square lattice is stabilized by two types of chemically specific interactions, each mediated by a square-well attraction of range $a/5$. One is a tetramer-stabilizing attraction of strength $-\epsilon_{\rm tet} \, \kt$, which binds together patches labeled `E' and `S' on neighboring monomers. The other is a dimer-stabilizing attraction of strength $-\epsilon_{\rm dim} \, \kt$, which binds together two patches labeled `L'. (These interactions were called $\epsilon_{\rm int}$ and $\epsilon_{\rm ext}$ in Refs.~\cite{Whitelam2010, Haxton2012slayer}.) The dimer-stabilizing attraction binds neighboring tetramers into a square lattice.

In addition to their specific interactions, monomers possess a nonspecific pairwise attraction of energy $-\epsilon_{\rm n} \, \kt$ if circumscribed rectangles of width $a+2 \Delta$ and length $la+2 \Delta$, where $\Delta = a/5$ (dashed rectangle in Fig.~\ref{model}) overlap. The nonspecific interaction allows formation of  the  dense amorphous protein clusters seen in AFM images of SbpA self-assembly~\cite{Chung2010}, and allows formation of macroscopic liquid phases implicated in protein self-assembly generally~\cite{ten1997enhancement,vekilovtwo,vekilov2012phase,lutsko2012theoretical}. In general, the nonspecific interaction mediates transitions from gas to liquid to solid, while specific interactions determine whether those phases consist of monomers, dimers, or tetramers.  

We solved the model's thermodynamics using the analytic mean-field theory described in Ref.~\cite{Haxton2012slayer}, whose predictions closely match the  results of direct equilibrium simulations. We used this theory to calculate the free energies of 8 possible bulk phases: fluids of monomers, dimers, trimers and tetramers; close-packed monomer, dimer and tetramer structures; and the target square lattice structure. We define the thermodynamic driving force for assembly, ${\cal F}$, as the free energy gap between the stable square lattice structure and the most stable fluid phase. In addition, we characterized the effective strength of pairwise interactions using the reduced second virial coefficient $B_2^{\star} \equiv B_2/B_2^{\rm hard \, core}$, where $B_2 = \left( 4\pi\right)^{-1} \int d{\bm r}_{ij}d\theta_{ij}\,  (1-{\rm e}^{-\beta U_{ij}})$. This integral runs over all angles and distances for which two monomers (called $i$ and $j$) interact; $U_{ij}$ is the monomer-monomer interaction energy. $B_2^{\rm hard \, core}$ is defined similarly, but for monomers whose attractive interactions have been switched off.

We determined self-assembly dynamics using virtual-move Monte Carlo simulations~\cite{Whitelam2007,whitelam2009role} (using the version of the algorithm described in Ref.~\cite{whitelam2011approximating}) of 1024 monomers at constant packing fraction, starting from well-mixed conditions. We employed the parameterization established in Ref.~\cite{Haxton2012slayer} to model overdamped motion in solution, ensuring that bound clusters of hydrodynamic radius $R$ diffuse according to the Stokes solution for spheres of radius $R$. Taking $a=3.9$ nm, $T=300$ K, and solution viscosity $\eta=1.00 \times 10^{-3}$ Pa s, each Monte Carlo (MC) cycle corresponds to 2.42 ns.  As in Refs.~\cite{Whitelam2007, Haxton2012slayer} we quantified square lattice assembly by recording the scaled yield $f=f_3(f_3/(f_3+f_2))^2$, where $f_3$ and $f_2$ are the fraction of monomers with two and three specific bonds satisfied, respectively.  The squared factor rewards crystal domains with large bulk-to-boundary ratios and penalizes systems with many small or stringy domains.  However, since it also penalizes systems with many free tetramers, we checked that our conclusions continue to hold when replacing $f$ with the unscaled yield $f_3$.  The supplemental figures display alternate versions of figures using this replacement.

\section{Results}

\begin{figure*}
\begin{center}
\includegraphics[width=\linewidth]{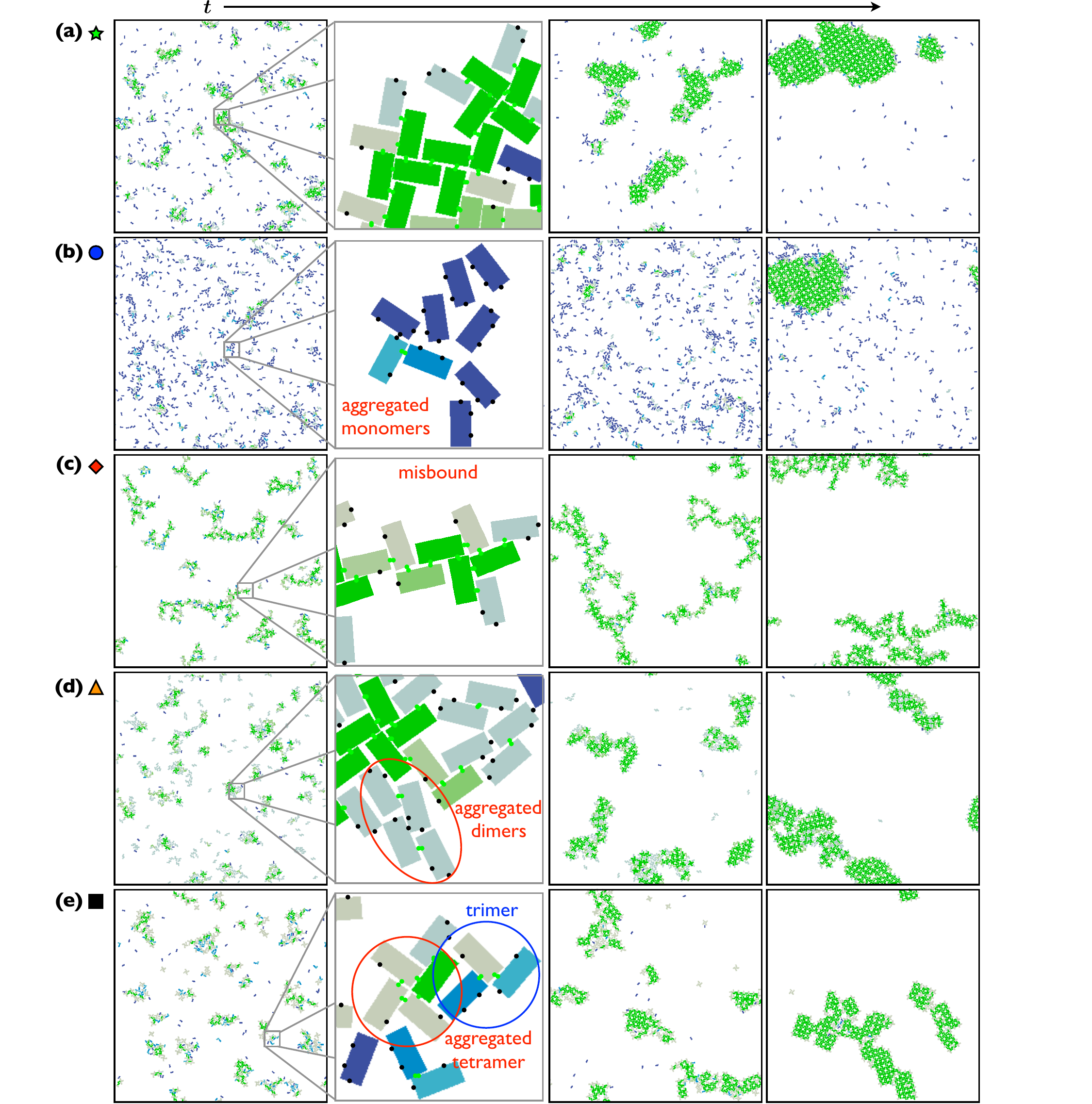}
\end{center}
\caption{\textit{Snapshots illustrate the diversity of self-assembly pathways seen throughout Fig.~\ref{example}.}  Time-ordered snapshots along assembly trajectories for \textbf{(a)} best assembly conditions and \textbf{(b-f)} four other points from Fig.~\ref{example}.  From left to right, the panels show the system after $10^5$ MC cycles (242 $\mu$s); important type(s) of microscopic environments seen in each trajectory; the system after $10^6$ MC cycles (2.42 ms); and the system after $10^7$ MC cycles (24.2 ms). Color code: crystalline monomers, with all three specific bonds satisfied, are green; monomers with both tetramer-stabilizing bonds satisfied are tan; monomers with its dimer-stabilizing bond satisfied are grey; monomers with only one tetramer-stabilizing bond satisfied are pale blue; and monomers with no specific bonds satisfied are dark blue. Symbols at left connect with their counterparts in Figs.~\ref{example} and \ref{timecourses}.}
\label{snapshots}
\end{figure*}

{\bf \em Assembly pathways and modes of kinetic trapping vary widely throughout parameter space.} Fig.~\ref{example} shows a color plot denoting the yield of the target square lattice structure obtained after long ($10^7$ MC cycles or 24.2 ms) dynamical simulations, overlaid on the equilibrium phase diagram. Grey lines on the phase diagram (derived from our analytic theory) show regions of stability of individual or coexisting phases, whose nature is illustrated by the associated simulation snapshots. The diagram is obtained by varying the specific interactions  $\epsilon_{\rm dim}$ and $\epsilon_{\rm tet}$ for one choice of nonspecific interaction strength, $\epsilon_{\rm n}=2$, and packing fraction, $\phi=0.1$. Yield is appreciable only in the upper right quadrant of the diagram, where both types of specific interactions are strong enough for the square lattice to be stable.  (While the other quadrants are not the focus of this paper, the phase behaviors found there are rich, comprising homogeneous fluid, liquid, and close-packed crystal phases of of monomers, dimers, and tetramers.)  Within the region of square lattice stability, assembly is generally good along the contour $\mathcal{F}= 2 \kt$ (corresponding to one of the design rules determined in our previous work~\cite{Haxton2012slayer}), and is best at the lower left corner of the contour. 

\begin{figure}[t]
\begin{center}
\includegraphics[width=\linewidth]{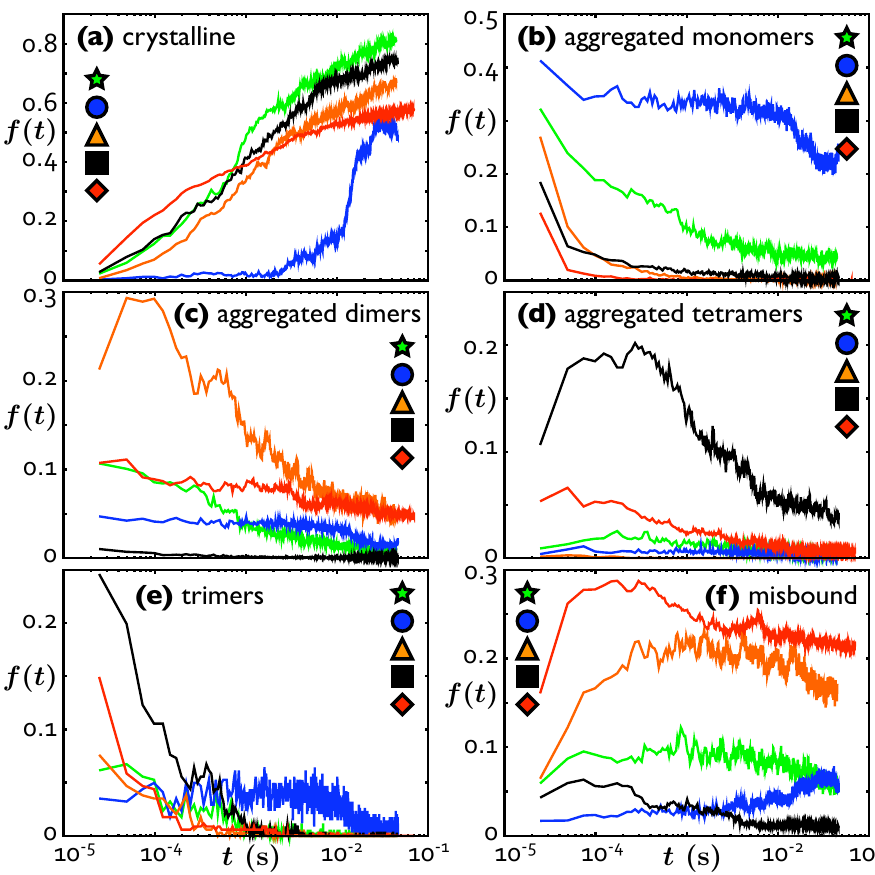}
\end{center}
\caption{\textit{Time evolution of microscopic environments reveals the variety of assembly pathways and modes of kinetic trapping seen throughout Fig.~\ref{example}.}  Each panel depicts the time evolution of a particular microscopic environment for the five points in phase space marked in Fig.~\ref{example} and illustrated in Fig.~\ref{snapshots}: \textbf{(a)} crystalline monomers having all three specific bonds satisfied; \textbf{(b-d)} aggregated monomers, dimers, and tetramers, respectively, defined as monomers with no specific bonds, one external bond, and both internal bonds satisfied, plus at least two additional nonspecific bonds; \textbf{(e)} trimers, defined as monomers belonging to a three-member cluster bound by specific interactions; and  \textbf{(f)}  `misbound' monomers, defined as monomers having a dimer-forming but only one tetramer-forming bond satisfied. Symbols at left connect with their counterparts in Figs.~\ref{example} and \ref{snapshots}.}
\label{timecourses}
\end{figure}

This localization can be understood by considering the dynamics of assembly pathways, which vary widely in character throughout this diagram. This variation is illustrated by snapshots in Fig.~\ref{snapshots} and time courses of several important microscopic environments in Fig.~\ref{timecourses}. The symbols -- green star, orange triangle etc. -- are common to Figs.~\ref{example}--\ref{timecourses}. 

When $\epsilon_{\rm dim}$ and $\epsilon_{\rm tet}$ are balanced in a roughly 2:1 ratio (e.g. at the green star on Fig.~\ref{example}), assembly occurs non-hierarchically, via the association of monomers. Yield is high in this regime where the thermodynamic driving force for assembly is between about 1 and 2 $\kt$ (note the green spots denoting high yield on Fig.~\ref{example}). Yield declines slightly in the nucleation regime near the phase boundary (e.g. at the blue circle) because of the lesser thermodynamic driving force for assembly, even though such assembly tends to result in only a single crystalline cluster. Similar observations have been made in other simulation models~\cite{grant2011analyzing,grant2012quantifying}.  Kinetic trapping occurs when the driving force for assembly is large (e.g. at the red diamond). Here, monomer-monomer contacts are too large in energy for mistakes in binding to be corrected before structures grow appreciably. These mistakes result in the accumulation of `misbound' monomers, those making a dimer-stabilizing bond but only one of two tetramer-stabilizing bonds, and the consequent assembly of gel-like structures with no counterpart on the thermodynamic phase diagram.  The accumulation of `misbound' monomers is depicted in the second panel of Fig.~\ref{snapshots} (c) and quantified by the red diamond trace in Fig.~\ref{timecourses}(f).

When $\epsilon_{\rm dim}$ is large (e.g. at the orange triangle), monomers rapidly form dimers which subsequently associate to form the target crystal structure. Even for an optimal thermodynamic driving force of 1 to 2 $k_{\rm B} T$, yield in this regime is slightly less than for the optimal non-hierarchical pathway, because dimers have a stronger tendency than monomers to associate nonspecifically: there are more ways two dimers can associate in a manner not commensurate with the target structure.  Further, the larger energy scales involved -- mediated by 4 particles instead of 2 -- mean that binding mistakes are less readily corrected, hampering crystal growth and preventing efficient coarsening of nuclei.  In particular, we find that dimers tend to aggregate nonspecifically with each other and coat the boundaries of growing crystal nuclei without reorienting efficiently to form specific bonds, reminiscent of the idea of surface ``blocking" by nonspecifically-bound proteins discussed in Ref.~\cite{Schmit2012}.  Such aggregation is depicted in the second panel of Fig.~\ref{snapshots} (d) and quantified by the orange diamond trace in Fig.~\ref{timecourses}(c).

Similar microscopics plagues the tetramer-dominated pathways, when $\epsilon_{\rm tet}$ is large (e.g. at the black square on Fig.~\ref{example}): tetramers also aggregate more readily than monomers. Note the peak associated with the black square trace in Fig.~\ref{timecourses}(d). In addition, the formation of trimers, which are not commensurate with the target structure, and the simultaneous depletion of the monomer pool, leads along tetramer-dominated pathways to an `incomplete building block' kinetic trap familiar from the study of virus capsids on the computer~\cite{hagan2006dynamic, Zlotnick1999, Zlotnick2003, Hagan2010, Hagan2011}. At the position marked by the black square on Fig.~\ref{example} the trimer acts more as a natural kinetic intermediate between dimers and tetramers than as a true `trap' (note the peak in the black square trace of Fig.~\ref{timecourses}(c), which diminishes at late times), but the associated trap becomes increasingly prominent as the energy scale associated with the tetramer-forming interaction increases.  The second panel of Fig.~\ref{snapshots} (e) depicts an aggregating tetramer and a trimer at early stages of assembly.\\

\begin{figure*}
\begin{center}
\includegraphics[width=\linewidth]{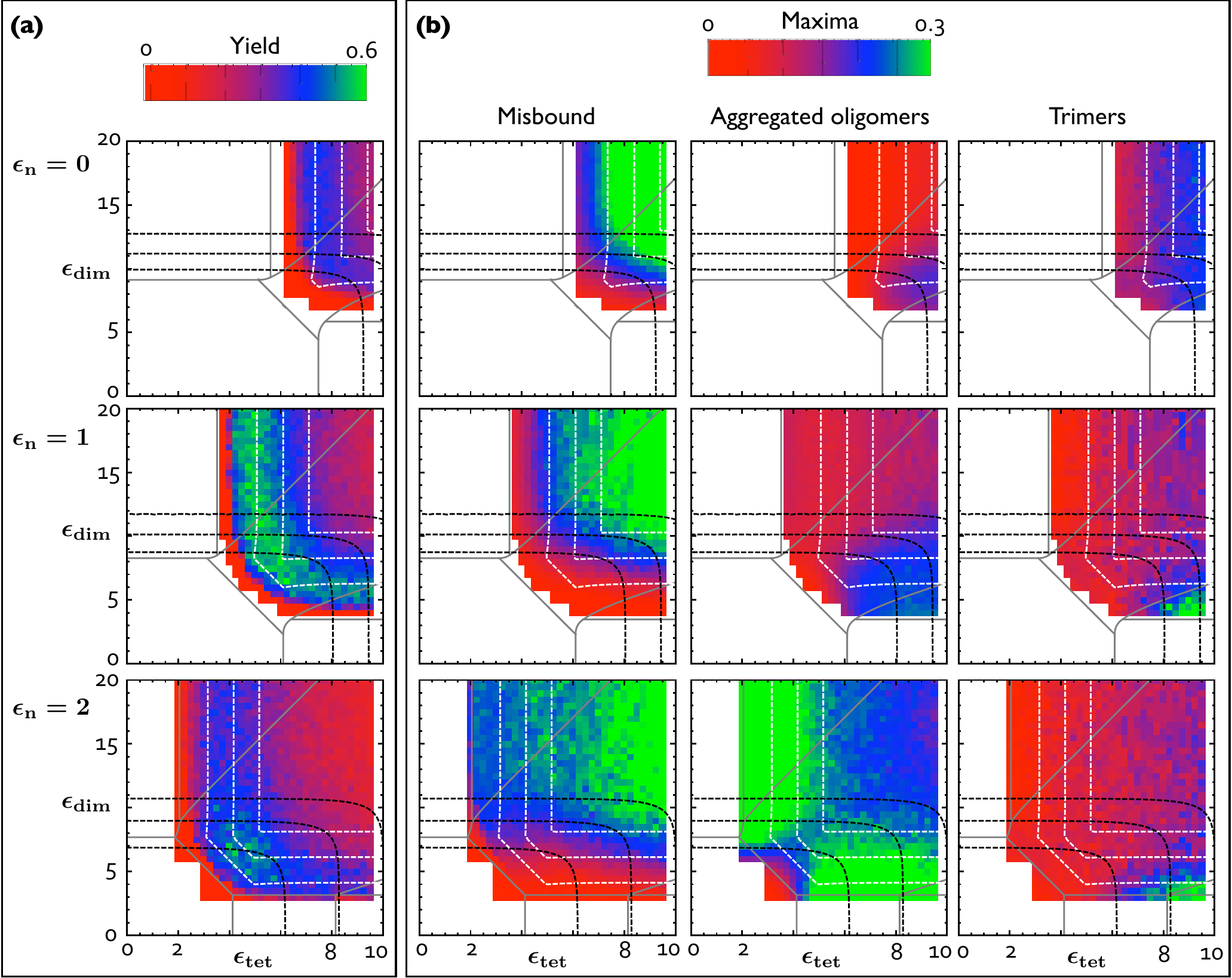}
\end{center}
\caption{{\em Weak nonspecific interaction is generally beneficial, but diminishes the effectiveness of hierarchical pathways relative to the direct one.} \textbf{(a)} Phase diagram and dynamic yield after $10^7$ MC cycles as a function of specific interaction strengths $\epsilon_{\rm tet}$ and $\epsilon_{\rm dim}$ for three nonspecific interaction strengths and a concentration $\phi=0.1$.  \textbf{(b)} Pathway diagrams (see text) illustrate that the regions of good assembly are bounded by regions exhibiting three types of kinetic trapping: misbinding, aggregation of oligomers (dimers or tetramers), and the formation of trimers. Weak nonspecific interaction (middle row) facilitates assembly, while stronger nonspecific interaction (bottom row) begins to suppress it. Furthermore, nonspecific modes of binding disfavor hierarchical pathways: moving from top to bottom, strengthening nonspecific interactions first disfavors assembly via tetramers (compare first and second rows), then disfavors assembly via dimers (compare second and third rows). }
\label{fixedenot}
\end{figure*}

\noindent
{\bf \em Weak nonspecific interaction is generally beneficial, but diminishes the relative effectiveness of hierarchical pathways.}  In Refs.~\cite{Whitelam2010,Haxton2012slayer} we found that weak nonspecific interactions facilitate assembly by allowing the strongest interactions (the specific ones) to be made weaker while maintaining an optimal thermodynamic driving force, thus enabling more efficient annealing of defects.  In Fig.~\ref{fixedenot}(a) we show that a similar trend is seen in the space of the two specific interactions: as nonspecific interaction $\epsilon_{\rm n}$ increases from 0 to $k_{\rm B}T$, best yield increases markedly.  However, if $\epsilon_{\rm n}$ is further increased, to $2k_{\rm B}T$ (the value used in Fig.~\ref{example}), yield generally decreases, consistent with our previous finding that density fluctuations associated with nonspecific association tend to conflict with the `symmetry fluctuations' required to form the square lattice~\cite{Haxton2012slayer}. Further, the nonspecific interaction acts to favor non-hierarchical pathways over hierarchical ones. For $\epsilon_{\rm n}=0$ peak yield (which is only moderate) does not differ appreciable from the non-hierarchical pathway sector (lower left of the assembly region) to the hierarchical pathway sectors (top left and bottom right portions of the assembly region). For $\epsilon_{\rm n}=1$, by contrast, the monomer and dimer pathways are noticably better than the tetrameter one. For still larger nonspecific interaction, $\epsilon_{\rm n}=2$, best yield is found in the region of non-hierarchical assembly.  Supplemental Fig. 1 shows that the same trends result when using the unscaled definition of yield, though the tetramer pathway is not noticeably worse than the dimer and monomer pathways at $\epsilon_{\rm n}=1$.  The relative efficiency of the direct pathway at large nonspecific interactions suggests that dimers and tetramers are more susceptible to nonspecific aggregation than are monomers, and that this aggregation proves deleterious to yield.

\begin{figure*}
\begin{center}
\includegraphics[width=\linewidth]{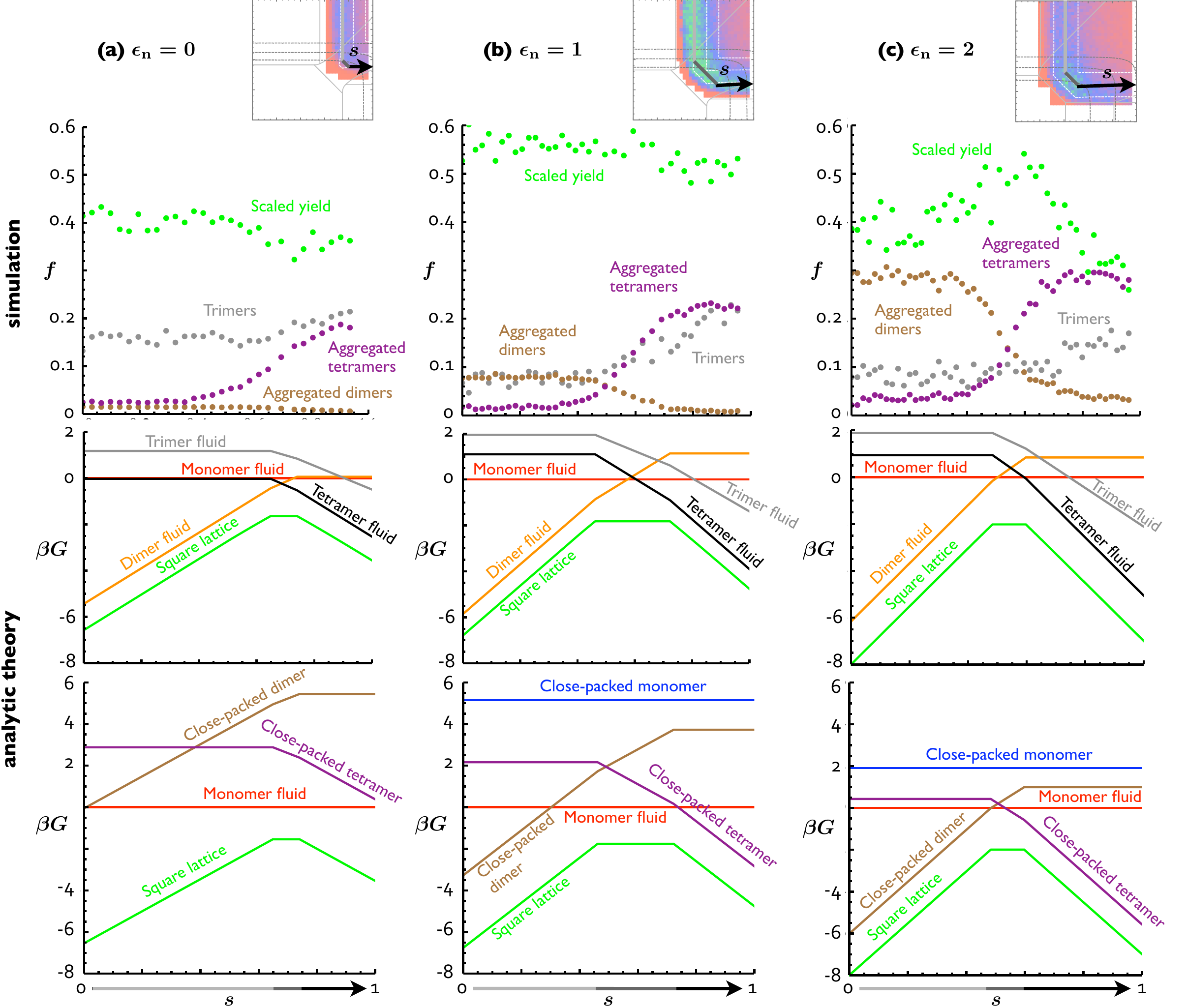}
\end{center}
\caption{ {\em Comparison of dynamics and thermodynamics reveals principles of good assembly.} Dynamic quantities (second row) respond to variations in the free energies of bulk phases in free energies (third and fourth rows), along paths, labeled $s$, of constant thermodynamic driving force (top row) through the phase diagrams of Fig.~\ref{fixedenot}. Shown are free energies of pure fluid and crystal phases; the free energies of equilibrium coexisting phases are equal or slightly lower. Hierarchical pathways owe their existence to dimer and tetrameter phases close to or lower than the parent phase in free energy; when these low-lying oligomer phases are close-packed (see right column, bottom panel), dynamic aggregation of oligomers leads to a reduction in yield (see right column, second panel). Non-hierarchical pathways resulting in high yield tend to occur when nothing lies between the monomer fluid and the target crystal in free energy.}
\label{freeenergies}
\end{figure*}

\begin{figure*}
\begin{center}
\includegraphics[width=\linewidth]{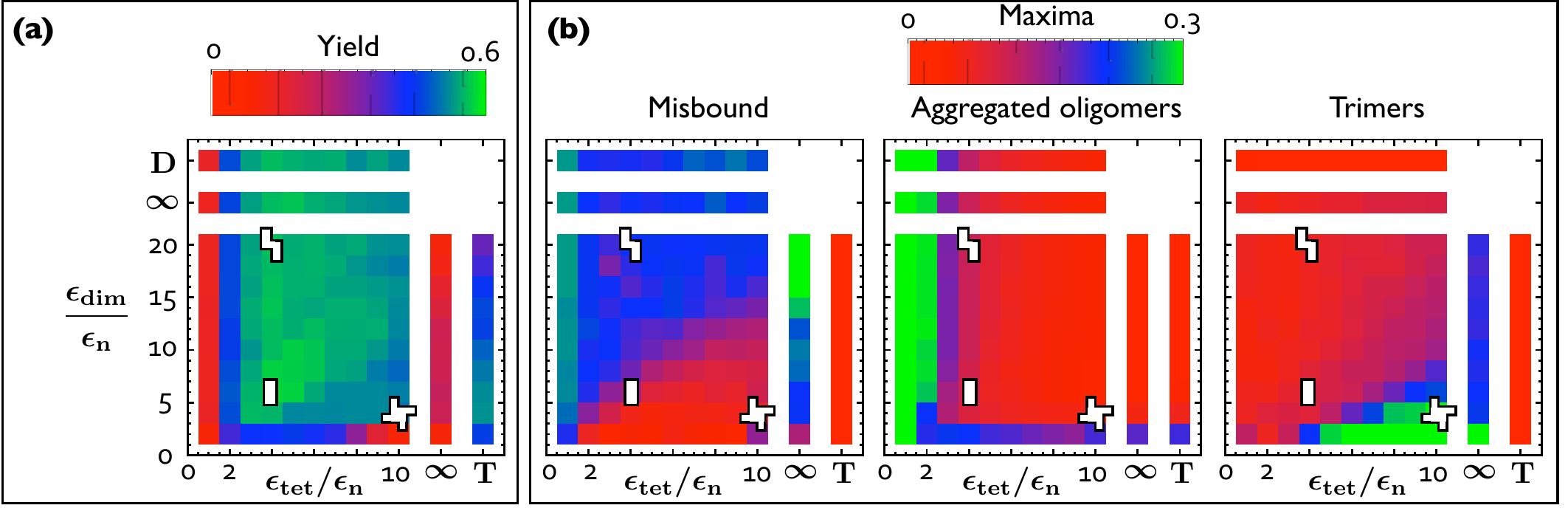}
\end{center}
\caption{\textit{Building blocks designed to assemble directly perform better than building blocks designed to assembly hierarchically, even under optimal conditions.}  \textbf{(a)} `Pathway design diagram' showing yield after $10^7$ MC cycles at $\phi=0.1$, for a wide range of building block designs defined by different interaction strength ratios $\epsilon_{\rm tet}/\epsilon_{\rm n}$ and $\epsilon_{\rm dim}/\epsilon_{\rm n}$. The absolute values of these parameters are chosen, at each point on the diagram, in order to optimize assembly; building blocks identified in this way are therefore the best-designed representatives of certain kinds of pathway. The symbol $\infty$ denotes unbreakable specific interactions, and the row (resp. column) marked `D' (resp. `T') correspond to simulations initialized with dimers (resp. tetramers) bound by unbreakable interactions. Comparison of yield and \textbf{(b)} pathways demonstrates that while oligomer aggregation has been largely eliminated, the trimer kinetic trap is always present when strong tetramer-forming bonds exist. Symbols show the points of intersection of this diagram with those shown in Fig.~\ref{phasediagrams}.}
\label{design}
\end{figure*}

This suggestion is confirmed by the `pathway diagrams'~\cite{Haxton2012slayer,hedges2011limit} shown in Fig.~\ref{fixedenot}(b). These denote the largest fraction of specific microscopic environments seen along trajectories at each point in phase space. These diagrams confirm that nonspecific aggregation of oligomers (dimers and tetramers) plagues much of phase space when the nonspecific interaction is large, while misbinding (the formation of only two out of three specific bonds) is the prevalent kinetic trap in the upper right sector of all diagrams, where one or both of the specific interactions are large. These two types of kinetic trapping bound the regions of optimal assembly, while the trimer `incomplete building block' trap poses an additional impediment to assembly in part of the tetramer-dominated phase space.\\

\noindent
{\bf \em Comparison of dynamics and thermodynamics reveals principles of good assembly.} Analysis of our analytic mean-field theory reveals the thermodynamic underpinning of the different dynamic pathways and modes of kinetic trapping seen in Fig.~\ref{fixedenot}. In Fig.~\ref{freeenergies} we show the three yield diagrams from Fig.~\ref{fixedenot} above a set of data collected along a contour of thermodynamic driving force ${\cal F}$, chosen to intersect the region of best yield in each case. Distance along the contour is labeled by the parameter $s$ (see top row of Fig.~\ref{freeenergies}). On the second row of Fig.~\ref{freeenergies} we show self-assembly yield and the largest fraction of specific microscopic environments seen in dynamic simulations perfomed at points along each contour. Below simulation results we show bulk free energies of 8 different phases from our analytic theory. As well as the monomer fluid phase (the parent phase) and the target square lattice structure, we show free energies of fluids of dimers, trimers and tetramers (third row of figure), and free energies of close-packed phases of dimers and tetramers (fourth row of figure). The arrow below each column corresponds to the contour drawn on each phase diagram at the top.  Supplemental Fig. 2 shows a similar version of Fig.~\ref{freeenergies} using the unscaled definition of yield.

Comparison of the dynamic and thermodynamic data yields two conclusions. First, hierarchical pathways owe their existence to phases (fluid and/or close-packed) of dimers and tetramers that lie between the parent phase and the target structure in free energy (or sit slightly above but close to the parent phase). Similar behavior is seen in the model of Ref.~\cite{PhysRevLett.102.118106}. This finding recalls Ostwald's rule of stages~\cite{ostwald1897studies,cardew1984kinetics,wolde1999homogeneous,threlfall2003structural,sear2005formation,hedges2011limit}, the heuristic that asserts that phases closest to the parent phase in free energy are visited before the stable phase. Second, good assembly occurs where neither close-packed oligomer nor trimer phases lie near or below the parent phase in free energy. The connection between the metastability of these phases and kinetic trapping is made clear by noting that microscopic environments in the second row have appreciable values where the associated phases (same colors) in the third and fourth rows lie near or below the monomer fluid phase.  This observation allows us to refine the second design rule governing best assembly in this model: do not allow dense liquid phases, extended close-packed solid phases, or fluids of incomplete building blocks to lie close in free energy to the target crystal structure.\\

\begin{figure*}
\begin{center}
\includegraphics[width=\linewidth]{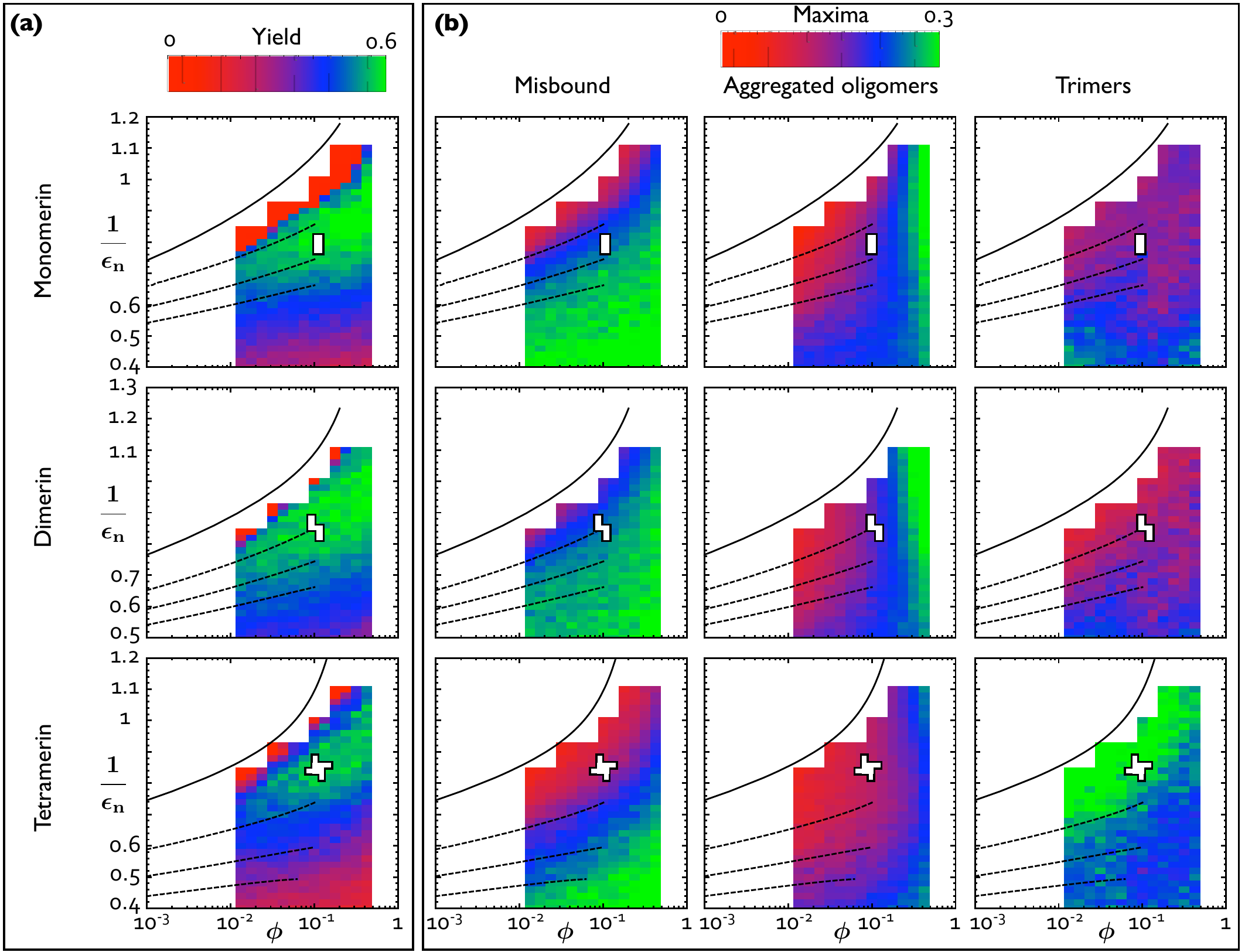}
\end{center}
\caption{{\em Apparently similar temperature-versus-packing fraction phase diagrams can conceal markedly different dynamics.} \textbf{(a)} Temperature-packing fraction phase diagrams for three building blocks representative of non-hierarchical, dimer-dominated, and tetramer-dominated assembly, from top to bottom. Symbols show the points of intersection of these diagram with that shown in Fig.~\ref{design}, equivalent to the temperature of peak yield for packing fraction $\phi=0.1$. The associated \textbf{(b)} pathway diagrams show that thermodynamic phase diagrams that look similar in this representation can harbor markedly different dynamics. In particular, tetramerin assembles via a trimer-rich intermediate, which acts as the dominant kinetic trap at high temperature and low packing fraction.}
\label{phasediagrams}
\end{figure*}

\noindent
{\bf \em The best-designed building blocks assemble directly rather than hierarchically.} Thus far we have focused on slices through parameter space in which the nonspecific interaction is held fixed and the specific interactions are varied.  In order to more thoroughly test whether hierarchical assembly pathways can outperform direct assembly pathways, we must design building blocks to avoid stabilizing unwanted phases while maintaining hierarchical assembly pathways.  Since the energetic part of the relative stability of target and close-packed phases are controlled by the ratios $\epsilon_{\rm tet}/\epsilon_{\rm n}$ and $\epsilon_{\rm dim}/\epsilon_{\rm n}$, tuning these ratios can alternately select hierarchical or direct pathways while ensuring that dense phases do not compete with the target square lattice.  Fig.~\ref{design} shows yield and pathway diagrams as functions of ratios $\epsilon_{\rm dim}/\epsilon_{\rm n}$ and $\epsilon_{\rm tet}/\epsilon_{\rm n}$, where packing fraction is held fixed at 10\% but the magnitude of the energies at each point is optimized for yield.  The rightmost and topmost portions of these diagrams represent unbreakable tetramer-forming and dimer-forming interactions, respectively, and denote simulations initialized from well-mixed initial configurations (labeled `$\infty$') or from gases of tetramers (labeled `T') or dimers (labeled `D').  Fig.~\ref{design} (a) shows that under optimal conditions, the non-hierarchical designs (near the diagonal $\epsilon_{\rm dim}/\epsilon_{\rm tet}=2$) assemble best, dimer-stabilizing designs (above the diagonal) assemble nearly as well (but not better), and tetramer-stabilizing designs (to the right of the diagonal) assemble much less well. Fig.~\ref{design} (b) confirms that this hierarchy is a result of kinetic trapping: while the aggregated oligomer trap has been all but eliminated, the trimer `incomplete building block' kinetic trap cannot be eliminated in the presence of strong tetramer-forming interactions. When this trap is circumvented artificially, by seeding simulations with a gas of tetramers (columns labeled `T'), yield in the tetramer-dominated sector is much improved.  Supplemental Fig. 3 shows that a similar conclusions result when using the unscaled definition of yield, though only strongly tetramer-stabilizing designs ($\epsilon_{\rm tet}/\epsilon_{\rm n}=\infty$ but not $\epsilon_{\rm tet}/\epsilon_{\rm n}\le10$) suppress yield.

Finally, we show in Fig.~\ref{phasediagrams} (and Supplemental Fig. 4) the behavior of three building block designs representative of non-hierarchical assembly, dimer-dominated assembly, and tetramer-dominated assembly, respectively, in a more conventional temperature-versus-packing fraction representation. Symbols show the points of intersection between these diagrams and the one shown in Fig.~\ref{design}. Comparison of phase diagrams, overlaid yield plots (panel a) and pathway diagrams (panel b) illustrates that three building blocks with similar-looking solubility curves and identical stable phases can display a wide variety of dynamic pathways to self-assembly, and exhibit markedly different kinds of kinetic trapping.

\section{Conclusion}

The self-assembly of the hierarchical structure studied here can occur via many dynamical pathways. These pathways can be non-hierarchical or hierarchical, can involve intermediates with counterparts on the equilibrium phase diagram -- such as dimers, tetramers, and dense fluid phases~\cite{Haxton2012slayer} -- or intermediates with no such counterpart, such as gel-like structures. Hierarchical assembly in our model is no better or worse than non-hierarchical assembly in the {\em absence} of nonspecific interactions between building blocks, but rapidly becomes worse than non-hierarchical assembly {\em when the possibility of nonspecific association exists}. In our model, hierarchical assembly is also vulnerable to the formation of incomplete higher-order building blocks (trimers) not commensurate with the target structure.  Models of virus capsid~\cite{hagan2006dynamic,Zlotnick1999,Zlotnick2003,Hagan2010,Hagan2011} and protein complex~\cite{PhysRevLett.102.118106} assembly suggest that the `incomplete building block' may be a general nuisance to the assembly of complicated structures, but that nature often mitigates this nuisance by favoring multi-step hierarchical assembly, such as the formation of capsomer hexamers as trimers of dimers~\cite{ChenB2011,Tsiang2012}.

In general, best assembly is achieved in our model when the dynamical pathway is simplest, comprising sequential addition of monomers to a single structure. The corresponding rules for design of the model's self-assembly pathways, gleaned from this work and our previous ones~\cite{Whitelam2010,Haxton2012slayer}, appear to stem from two simple thermodynamic prescriptions: 1) make the free energy gap between the target phase and the most stable fluid phase of order $k_{\rm B} T$, and 2) ensure that no other dense phase (liquid of close-packed solid of monomers or oligomers)  or incomplete-building-block phase falls within this gap.  Achieving this prescription requires tuning the three microscopic interactions possessed by the model, and, in particular, making the nonspecific interaction weak. The connection between underlying thermodynamics and self-assembly dynamics we have made helps us place the model in a broader context: its assembly is similar in certain regards to that of virus capsids, but mostly dissimilar to the assembly of close-packed crystals. Continued classification of the assembly pathways of numerous types of structure~\cite{hsu2010theoretical,sciortino2007self,de2006dynamics,wilber2009self,Jankowski2012} should eventually reveal whether apparently dissimilar systems can nonetheless be placed in similar categories of behaviors.







\footnotesize{

\providecommand*{\mcitethebibliography}{\thebibliography}
\csname @ifundefined\endcsname{endmcitethebibliography}
{\let\endmcitethebibliography\endthebibliography}{}

}


\begin{mcitethebibliography}{57}
\providecommand*{\natexlab}[1]{#1}
\providecommand*{\mciteSetBstSublistMode}[1]{}
\providecommand*{\mciteSetBstMaxWidthForm}[2]{}
\providecommand*{\mciteBstWouldAddEndPuncttrue}
  {\def\EndOfBibitem{\unskip.}}
\providecommand*{\mciteBstWouldAddEndPunctfalse}
  {\let\EndOfBibitem\relax}
\providecommand*{\mciteSetBstMidEndSepPunct}[3]{}
\providecommand*{\mciteSetBstSublistLabelBeginEnd}[3]{}
\providecommand*{\EndOfBibitem}{}
\mciteSetBstSublistMode{f}
\mciteSetBstMaxWidthForm{subitem}
{(\emph{\alph{mcitesubitemcount}})}
\mciteSetBstSublistLabelBeginEnd{\mcitemaxwidthsubitemform\space}
{\relax}{\relax}

\bibitem[Vekilov()]{vekilovtwo}
P.~G. Vekilov, \emph{Kinetics and Thermodynamics of Multistep Nucleation and
  Self-Assembly in Nanoscale Materials},  79--109\relax
\mciteBstWouldAddEndPuncttrue
\mciteSetBstMidEndSepPunct{\mcitedefaultmidpunct}
{\mcitedefaultendpunct}{\mcitedefaultseppunct}\relax
\EndOfBibitem
\bibitem[Becker and D{\"o}ring(1935)]{becker1935kinetische}
R.~Becker and W.~D{\"o}ring, \emph{Annalen der Physik}, 1935, \textbf{416},
  719--752\relax
\mciteBstWouldAddEndPuncttrue
\mciteSetBstMidEndSepPunct{\mcitedefaultmidpunct}
{\mcitedefaultendpunct}{\mcitedefaultseppunct}\relax
\EndOfBibitem
\bibitem[Volmer and Weber(1926)]{volmer1926keimbildung}
M.~Volmer and A.~Weber, \emph{Z. Phys. Chem}, 1926, \textbf{119},
  277--301\relax
\mciteBstWouldAddEndPuncttrue
\mciteSetBstMidEndSepPunct{\mcitedefaultmidpunct}
{\mcitedefaultendpunct}{\mcitedefaultseppunct}\relax
\EndOfBibitem
\bibitem[Whitelam \emph{et~al.}(2009)Whitelam, Feng, Hagan, and
  Geissler]{whitelam2009role}
S.~Whitelam, E.~H. Feng, M.~F. Hagan and P.~L. Geissler, \emph{Soft Matter},
  2009, \textbf{5}, 1251--1262\relax
\mciteBstWouldAddEndPuncttrue
\mciteSetBstMidEndSepPunct{\mcitedefaultmidpunct}
{\mcitedefaultendpunct}{\mcitedefaultseppunct}\relax
\EndOfBibitem
\bibitem[Villar \emph{et~al.}(2009)Villar, Wilber, Williamson, Thiara, Doye,
  Louis, Jochum, Lewis, and Levy]{PhysRevLett.102.118106}
G.~Villar, A.~W. Wilber, A.~J. Williamson, P.~Thiara, J.~P.~K. Doye, A.~A.
  Louis, M.~N. Jochum, A.~C.~F. Lewis and E.~D. Levy, \emph{Phys. Rev. Lett.},
  2009, \textbf{102}, 118106\relax
\mciteBstWouldAddEndPuncttrue
\mciteSetBstMidEndSepPunct{\mcitedefaultmidpunct}
{\mcitedefaultendpunct}{\mcitedefaultseppunct}\relax
\EndOfBibitem
\bibitem[Bray(1994)]{bray1994theory}
A.~J. Bray, \emph{Advances in Physics}, 1994, \textbf{43}, 357--459\relax
\mciteBstWouldAddEndPuncttrue
\mciteSetBstMidEndSepPunct{\mcitedefaultmidpunct}
{\mcitedefaultendpunct}{\mcitedefaultseppunct}\relax
\EndOfBibitem
\bibitem[Cates and Evans(2000)]{cates2000soft}
M.~E. Cates and M.~R. Evans, \emph{Soft and Fragile Matter: Nonequilibrium
  Dynamics, Metastability and Flow (PBK)}, Taylor \& Francis, 2000,
  vol.~53\relax
\mciteBstWouldAddEndPuncttrue
\mciteSetBstMidEndSepPunct{\mcitedefaultmidpunct}
{\mcitedefaultendpunct}{\mcitedefaultseppunct}\relax
\EndOfBibitem
\bibitem[Fallas \emph{et~al.}(2010)Fallas, O'Leary, and Hartgerink]{Fallas2010}
J.~A. Fallas, L.~E.~R. O'Leary and J.~D. Hartgerink, \emph{Chem. Soc. Rev.},
  2010, \textbf{39}, 3510\relax
\mciteBstWouldAddEndPuncttrue
\mciteSetBstMidEndSepPunct{\mcitedefaultmidpunct}
{\mcitedefaultendpunct}{\mcitedefaultseppunct}\relax
\EndOfBibitem
\bibitem[Fang \emph{et~al.}(2011)Fang, Conway, Margolis, Simmer, and
  Beniash]{Fang2011}
P.-A. Fang, J.~F. Conway, H.~C. Margolis, J.~P. Simmer and E.~Beniash,
  \emph{Proc. Natl. Acad. Sci. U.S.A.}, 2011, \textbf{108}, 14097\relax
\mciteBstWouldAddEndPuncttrue
\mciteSetBstMidEndSepPunct{\mcitedefaultmidpunct}
{\mcitedefaultendpunct}{\mcitedefaultseppunct}\relax
\EndOfBibitem
\bibitem[Zlotnick and Mukhopadhyay(2011)]{Zlotnick2011}
A.~Zlotnick and S.~Mukhopadhyay, \emph{Trends Microbiol.}, 2011, \textbf{19},
  14\relax
\mciteBstWouldAddEndPuncttrue
\mciteSetBstMidEndSepPunct{\mcitedefaultmidpunct}
{\mcitedefaultendpunct}{\mcitedefaultseppunct}\relax
\EndOfBibitem
\bibitem[Herrmann and Aebi(2004)]{Herrmann2004}
H.~Herrmann and U.~Aebi, \emph{Annu. Rev. Biochem.}, 2004, \textbf{73},
  749\relax
\mciteBstWouldAddEndPuncttrue
\mciteSetBstMidEndSepPunct{\mcitedefaultmidpunct}
{\mcitedefaultendpunct}{\mcitedefaultseppunct}\relax
\EndOfBibitem
\bibitem[Mann(2001)]{mann2001biomineralization}
S.~Mann, \emph{Biomineralization: principles and concepts in bioinorganic
  materials chemistry}, Oxford University Press, USA, 2001, vol.~5\relax
\mciteBstWouldAddEndPuncttrue
\mciteSetBstMidEndSepPunct{\mcitedefaultmidpunct}
{\mcitedefaultendpunct}{\mcitedefaultseppunct}\relax
\EndOfBibitem
\bibitem[Dove \emph{et~al.}(2003)Dove, De~Yoreo, and
  Weiner]{dove2003biomineralization}
P.~M. Dove, J.~De~Yoreo and S.~Weiner, \emph{Biomineralization}, Mineralogical
  Society of America, 2003\relax
\mciteBstWouldAddEndPuncttrue
\mciteSetBstMidEndSepPunct{\mcitedefaultmidpunct}
{\mcitedefaultendpunct}{\mcitedefaultseppunct}\relax
\EndOfBibitem
\bibitem[Fratzl and Weinkamer(2007)]{fratzl2007nature}
P.~Fratzl and R.~Weinkamer, \emph{Progress in Materials Science}, 2007,
  \textbf{52}, 1263--1334\relax
\mciteBstWouldAddEndPuncttrue
\mciteSetBstMidEndSepPunct{\mcitedefaultmidpunct}
{\mcitedefaultendpunct}{\mcitedefaultseppunct}\relax
\EndOfBibitem
\bibitem[Li \emph{et~al.}(2012)Li, Nielsen, Lee, Frandsen, Banfield, and
  De~Yoreo]{li2012direction}
D.~Li, M.~H. Nielsen, J.~R.~I. Lee, C.~Frandsen, J.~F. Banfield and J.~J.
  De~Yoreo, \emph{Science}, 2012, \textbf{336}, 1014--1018\relax
\mciteBstWouldAddEndPuncttrue
\mciteSetBstMidEndSepPunct{\mcitedefaultmidpunct}
{\mcitedefaultendpunct}{\mcitedefaultseppunct}\relax
\EndOfBibitem
\bibitem[Liao \emph{et~al.}(2012)Liao, Cui, Whitelam, and Zheng]{liao2012real}
H.~G. Liao, L.~Cui, S.~Whitelam and H.~Zheng, \emph{science}, 2012,
  \textbf{336}, 1011--1014\relax
\mciteBstWouldAddEndPuncttrue
\mciteSetBstMidEndSepPunct{\mcitedefaultmidpunct}
{\mcitedefaultendpunct}{\mcitedefaultseppunct}\relax
\EndOfBibitem
\bibitem[Murnen \emph{et~al.}(2010)Murnen, Rosales, Jaworsk, Segalman, and
  Zuckermann]{Murnen2010}
H.~K. Murnen, A.~M. Rosales, J.~N. Jaworsk, R.~A. Segalman and R.~N.
  Zuckermann, \emph{J. Am. Chem. Soc.}, 2010, \textbf{132}, 16112\relax
\mciteBstWouldAddEndPuncttrue
\mciteSetBstMidEndSepPunct{\mcitedefaultmidpunct}
{\mcitedefaultendpunct}{\mcitedefaultseppunct}\relax
\EndOfBibitem
\bibitem[Miszta \emph{et~al.}(2011)Miszta, de~Graaf, Bertoni, Dorfs, Brescia,
  Marras, Ceseracciu, Cingolani, van Roij, Dijkstra, and Manna]{Miszta2011}
K.~Miszta, J.~de~Graaf, G.~Bertoni, D.~Dorfs, R.~Brescia, S.~Marras,
  L.~Ceseracciu, R.~Cingolani, R.~van Roij, M.~Dijkstra and L.~Manna,
  \emph{Nature Mater.}, 2011, \textbf{10}, 872\relax
\mciteBstWouldAddEndPuncttrue
\mciteSetBstMidEndSepPunct{\mcitedefaultmidpunct}
{\mcitedefaultendpunct}{\mcitedefaultseppunct}\relax
\EndOfBibitem
\bibitem[Tao \emph{et~al.}(2012)Tao, Shen, Yang, Han, Huang, Li, Chu, and
  Xie]{Tao2012}
Y.~Tao, Y.~Shen, L.~Yang, B.~Han, F.~Huang, S.~Li, Z.~Chu and A.~Xie,
  \emph{Nanoscale}, 2012, \textbf{4}, 3729\relax
\mciteBstWouldAddEndPuncttrue
\mciteSetBstMidEndSepPunct{\mcitedefaultmidpunct}
{\mcitedefaultendpunct}{\mcitedefaultseppunct}\relax
\EndOfBibitem
\bibitem[Rehm \emph{et~al.}(2012)Rehm, Gr\:{o}hn, and Schmuck]{Rehm2012}
T.~H. Rehm, F.~Gr\:{o}hn and C.~Schmuck, \emph{Soft Matter}, 2012, \textbf{8},
  3154\relax
\mciteBstWouldAddEndPuncttrue
\mciteSetBstMidEndSepPunct{\mcitedefaultmidpunct}
{\mcitedefaultendpunct}{\mcitedefaultseppunct}\relax
\EndOfBibitem
\bibitem[Pum \emph{et~al.}(1991)Pum, Messner, and Sleytr]{pum1991role}
D.~Pum, P.~Messner and U.~B. Sleytr, \emph{Journal of bacteriology}, 1991,
  \textbf{173}, 6865--6873\relax
\mciteBstWouldAddEndPuncttrue
\mciteSetBstMidEndSepPunct{\mcitedefaultmidpunct}
{\mcitedefaultendpunct}{\mcitedefaultseppunct}\relax
\EndOfBibitem
\bibitem[Messner and Sleytr(1992)]{messner1992crystalline}
P.~Messner and U.~B. Sleytr, \emph{Advances in microbial physiology}, 1992,
  \textbf{33}, 213\relax
\mciteBstWouldAddEndPuncttrue
\mciteSetBstMidEndSepPunct{\mcitedefaultmidpunct}
{\mcitedefaultendpunct}{\mcitedefaultseppunct}\relax
\EndOfBibitem
\bibitem[Sleytr(1997)]{sleytr1997baa}
U.~B. Sleytr, \emph{FEMS Microbiology Reviews}, 1997, \textbf{20}, 5--12\relax
\mciteBstWouldAddEndPuncttrue
\mciteSetBstMidEndSepPunct{\mcitedefaultmidpunct}
{\mcitedefaultendpunct}{\mcitedefaultseppunct}\relax
\EndOfBibitem
\bibitem[Chung \emph{et~al.}(2010)Chung, Shin, Bertozzi, and
  De~Yoreo]{chung2010self}
S.~Chung, S.~H. Shin, C.~R. Bertozzi and J.~J. De~Yoreo, \emph{Proceedings of
  the National Academy of Sciences}, 2010, \textbf{107}, 16536--16541\relax
\mciteBstWouldAddEndPuncttrue
\mciteSetBstMidEndSepPunct{\mcitedefaultmidpunct}
{\mcitedefaultendpunct}{\mcitedefaultseppunct}\relax
\EndOfBibitem
\bibitem[Haxton and Whitelam(2012)]{Haxton2012slayer}
T.~K. Haxton and S.~Whitelam, \emph{Soft Matter}, 2012, \textbf{8}, 3558\relax
\mciteBstWouldAddEndPuncttrue
\mciteSetBstMidEndSepPunct{\mcitedefaultmidpunct}
{\mcitedefaultendpunct}{\mcitedefaultseppunct}\relax
\EndOfBibitem
\bibitem[Sweeney \emph{et~al.}(2008)Sweeney, Zhang, and
  Schwartz]{sweeney2008exploring}
B.~Sweeney, T.~Zhang and R.~Schwartz, \emph{Biophysical journal}, 2008,
  \textbf{94}, 772--783\relax
\mciteBstWouldAddEndPuncttrue
\mciteSetBstMidEndSepPunct{\mcitedefaultmidpunct}
{\mcitedefaultendpunct}{\mcitedefaultseppunct}\relax
\EndOfBibitem
\bibitem[Hagan and Chandler(2006)]{hagan2006dynamic}
M.~F. Hagan and D.~Chandler, \emph{Biophysical journal}, 2006, \textbf{91},
  42--54\relax
\mciteBstWouldAddEndPuncttrue
\mciteSetBstMidEndSepPunct{\mcitedefaultmidpunct}
{\mcitedefaultendpunct}{\mcitedefaultseppunct}\relax
\EndOfBibitem
\bibitem[Rapaport(2008)]{rapaport2008role}
D.~C. Rapaport, \emph{Physical Review Letters}, 2008, \textbf{101},
  186101\relax
\mciteBstWouldAddEndPuncttrue
\mciteSetBstMidEndSepPunct{\mcitedefaultmidpunct}
{\mcitedefaultendpunct}{\mcitedefaultseppunct}\relax
\EndOfBibitem
\bibitem[Nguyen \emph{et~al.}(2009)Nguyen, Reddy, and
  Brooks~Iii]{nguyen2009invariant}
H.~D. Nguyen, V.~S. Reddy and C.~L. Brooks~Iii, \emph{Journal of the American
  Chemical Society}, 2009, \textbf{131}, 2606--2614\relax
\mciteBstWouldAddEndPuncttrue
\mciteSetBstMidEndSepPunct{\mcitedefaultmidpunct}
{\mcitedefaultendpunct}{\mcitedefaultseppunct}\relax
\EndOfBibitem
\bibitem[ten Wolde and Frenkel(1997)]{ten1997enhancement}
P.~R. ten Wolde and D.~Frenkel, \emph{Science}, 1997, \textbf{277},
  1975--1978\relax
\mciteBstWouldAddEndPuncttrue
\mciteSetBstMidEndSepPunct{\mcitedefaultmidpunct}
{\mcitedefaultendpunct}{\mcitedefaultseppunct}\relax
\EndOfBibitem
\bibitem[Xu \emph{et~al.}(2012)Xu, Buldyrev, Stanley, and
  Franzese]{xu2012homogeneous}
L.~Xu, S.~V. Buldyrev, H.~E. Stanley and G.~Franzese, \emph{Physical Review
  Letters}, 2012, \textbf{109}, 95702\relax
\mciteBstWouldAddEndPuncttrue
\mciteSetBstMidEndSepPunct{\mcitedefaultmidpunct}
{\mcitedefaultendpunct}{\mcitedefaultseppunct}\relax
\EndOfBibitem
\bibitem[Zlotnick \emph{et~al.}(1999)Zlotnick, Johnson, Wingfield, and
  Stahl]{Zlotnick1999}
A.~Zlotnick, J.~M. Johnson, P.~W. Wingfield and S.~Stahl, \emph{Biochemistry},
  1999, \textbf{38}, 14644\relax
\mciteBstWouldAddEndPuncttrue
\mciteSetBstMidEndSepPunct{\mcitedefaultmidpunct}
{\mcitedefaultendpunct}{\mcitedefaultseppunct}\relax
\EndOfBibitem
\bibitem[Zlotnick(2003)]{Zlotnick2003}
A.~Zlotnick, \emph{Virology}, 2003, \textbf{315}, 269\relax
\mciteBstWouldAddEndPuncttrue
\mciteSetBstMidEndSepPunct{\mcitedefaultmidpunct}
{\mcitedefaultendpunct}{\mcitedefaultseppunct}\relax
\EndOfBibitem
\bibitem[Hagan and Elrad(2010)]{Hagan2010}
M.~F. Hagan and O.~M. Elrad, \emph{Biophys. J.}, 2010, \textbf{98}, 1065\relax
\mciteBstWouldAddEndPuncttrue
\mciteSetBstMidEndSepPunct{\mcitedefaultmidpunct}
{\mcitedefaultendpunct}{\mcitedefaultseppunct}\relax
\EndOfBibitem
\bibitem[Hagan \emph{et~al.}(2011)Hagan, Elrad, and Jack]{Hagan2011}
M.~F. Hagan, O.~M. Elrad and R.~L. Jack, \emph{J. Chem. Phys.}, 2011,
  \textbf{135}, 104115\relax
\mciteBstWouldAddEndPuncttrue
\mciteSetBstMidEndSepPunct{\mcitedefaultmidpunct}
{\mcitedefaultendpunct}{\mcitedefaultseppunct}\relax
\EndOfBibitem
\bibitem[Whitelam(2010)]{Whitelam2010}
S.~Whitelam, \emph{Phys. Rev. Lett.}, 2010, \textbf{105}, 088102\relax
\mciteBstWouldAddEndPuncttrue
\mciteSetBstMidEndSepPunct{\mcitedefaultmidpunct}
{\mcitedefaultendpunct}{\mcitedefaultseppunct}\relax
\EndOfBibitem
\bibitem[Chung \emph{et~al.}(2010)Chung, Shin, Bertozzi, and
  De~Yoreo]{Chung2010}
S.~Chung, S.-H. Shin, C.~R. Bertozzi and J.~J. De~Yoreo, \emph{Proc. Natl.
  Acad. Sci. U.S.A.}, 2010, \textbf{107}, 16536\relax
\mciteBstWouldAddEndPuncttrue
\mciteSetBstMidEndSepPunct{\mcitedefaultmidpunct}
{\mcitedefaultendpunct}{\mcitedefaultseppunct}\relax
\EndOfBibitem
\bibitem[Vekilov(2012)]{vekilov2012phase}
P.~G. Vekilov, \emph{Journal of Physics: Condensed Matter}, 2012, \textbf{24},
  193101\relax
\mciteBstWouldAddEndPuncttrue
\mciteSetBstMidEndSepPunct{\mcitedefaultmidpunct}
{\mcitedefaultendpunct}{\mcitedefaultseppunct}\relax
\EndOfBibitem
\bibitem[Lutsko(2012)]{lutsko2012theoretical}
J.~Lutsko, \emph{Bulletin of the American Physical Society}, 2012, \textbf{57},
  year\relax
\mciteBstWouldAddEndPuncttrue
\mciteSetBstMidEndSepPunct{\mcitedefaultmidpunct}
{\mcitedefaultendpunct}{\mcitedefaultseppunct}\relax
\EndOfBibitem
\bibitem[Whitelam and Geissler(2007)]{Whitelam2007}
S.~Whitelam and P.~L. Geissler, \emph{J. Chem. Phys.}, 2007, \textbf{127},
  154101\relax
\mciteBstWouldAddEndPuncttrue
\mciteSetBstMidEndSepPunct{\mcitedefaultmidpunct}
{\mcitedefaultendpunct}{\mcitedefaultseppunct}\relax
\EndOfBibitem
\bibitem[Whitelam(2011)]{whitelam2011approximating}
S.~Whitelam, \emph{Molecular Simulation}, 2011, \textbf{37}, 606\relax
\mciteBstWouldAddEndPuncttrue
\mciteSetBstMidEndSepPunct{\mcitedefaultmidpunct}
{\mcitedefaultendpunct}{\mcitedefaultseppunct}\relax
\EndOfBibitem
\bibitem[Grant \emph{et~al.}(2011)Grant, Jack, and
  Whitelam]{grant2011analyzing}
J.~Grant, R.~L. Jack and S.~Whitelam, \emph{The Journal of Chemical Physics},
  2011, \textbf{135}, 214505--214505\relax
\mciteBstWouldAddEndPuncttrue
\mciteSetBstMidEndSepPunct{\mcitedefaultmidpunct}
{\mcitedefaultendpunct}{\mcitedefaultseppunct}\relax
\EndOfBibitem
\bibitem[Grant and Jack(2012)]{grant2012quantifying}
J.~Grant and R.~L. Jack, \emph{Physical Review E}, 2012, \textbf{85},
  021112\relax
\mciteBstWouldAddEndPuncttrue
\mciteSetBstMidEndSepPunct{\mcitedefaultmidpunct}
{\mcitedefaultendpunct}{\mcitedefaultseppunct}\relax
\EndOfBibitem
\bibitem[Schmit and Dill(2012)]{Schmit2012}
J.~D. Schmit and K.~Dill, \emph{J. Am. Chem. Soc.}, 2012, \textbf{134},
  3934\relax
\mciteBstWouldAddEndPuncttrue
\mciteSetBstMidEndSepPunct{\mcitedefaultmidpunct}
{\mcitedefaultendpunct}{\mcitedefaultseppunct}\relax
\EndOfBibitem
\bibitem[Hedges and Whitelam(2011)]{hedges2011limit}
L.~O. Hedges and S.~Whitelam, \emph{J. Chem. Phys.}, 2011, \textbf{135},
  164902\relax
\mciteBstWouldAddEndPuncttrue
\mciteSetBstMidEndSepPunct{\mcitedefaultmidpunct}
{\mcitedefaultendpunct}{\mcitedefaultseppunct}\relax
\EndOfBibitem
\bibitem[Ostwald(1897)]{ostwald1897studies}
W.~Z. Ostwald, \emph{Z. Phys. Chem.}, 1897, \textbf{22}, year\relax
\mciteBstWouldAddEndPuncttrue
\mciteSetBstMidEndSepPunct{\mcitedefaultmidpunct}
{\mcitedefaultendpunct}{\mcitedefaultseppunct}\relax
\EndOfBibitem
\bibitem[Cardew \emph{et~al.}(1984)Cardew, Davey, and
  Ruddick]{cardew1984kinetics}
P.~T. Cardew, R.~J. Davey and A.~J. Ruddick, \emph{Journal of the Chemical
  Society, Faraday Transactions 2}, 1984, \textbf{80}, 659--668\relax
\mciteBstWouldAddEndPuncttrue
\mciteSetBstMidEndSepPunct{\mcitedefaultmidpunct}
{\mcitedefaultendpunct}{\mcitedefaultseppunct}\relax
\EndOfBibitem
\bibitem[Wolde and Frenkel(1999)]{wolde1999homogeneous}
P.~R. Wolde and D.~Frenkel, \emph{Physical Chemistry Chemical Physics}, 1999,
  \textbf{1}, 2191--2196\relax
\mciteBstWouldAddEndPuncttrue
\mciteSetBstMidEndSepPunct{\mcitedefaultmidpunct}
{\mcitedefaultendpunct}{\mcitedefaultseppunct}\relax
\EndOfBibitem
\bibitem[Threlfall(2003)]{threlfall2003structural}
T.~Threlfall, \emph{Organic Process Research \& Development}, 2003, \textbf{7},
  1017--1027\relax
\mciteBstWouldAddEndPuncttrue
\mciteSetBstMidEndSepPunct{\mcitedefaultmidpunct}
{\mcitedefaultendpunct}{\mcitedefaultseppunct}\relax
\EndOfBibitem
\bibitem[Sear(2005)]{sear2005formation}
R.~P. Sear, \emph{Journal of Physics: Condensed Matter}, 2005, \textbf{17},
  3997\relax
\mciteBstWouldAddEndPuncttrue
\mciteSetBstMidEndSepPunct{\mcitedefaultmidpunct}
{\mcitedefaultendpunct}{\mcitedefaultseppunct}\relax
\EndOfBibitem
\bibitem[Chen and Tycko(2011)]{ChenB2011}
B.~Chen and R.~Tycko, \emph{Biophys. J.}, 2011, \textbf{100}, 3035\relax
\mciteBstWouldAddEndPuncttrue
\mciteSetBstMidEndSepPunct{\mcitedefaultmidpunct}
{\mcitedefaultendpunct}{\mcitedefaultseppunct}\relax
\EndOfBibitem
\bibitem[Tsiang \emph{et~al.}(2012)Tsiang, Niedziela-Majka, Hung, Jin, Hu,
  Yant, Samuel, Liu, and Sakowicz]{Tsiang2012}
M.~Tsiang, A.~Niedziela-Majka, M.~Hung, D.~Jin, E.~Hu, S.~Yant, D.~Samuel,
  X.~Liu and R.~Sakowicz, \emph{Biochemistry}, 2012, \textbf{51}, 4416\relax
\mciteBstWouldAddEndPuncttrue
\mciteSetBstMidEndSepPunct{\mcitedefaultmidpunct}
{\mcitedefaultendpunct}{\mcitedefaultseppunct}\relax
\EndOfBibitem
\bibitem[Hsu \emph{et~al.}(2010)Hsu, Sciortino, and Starr]{hsu2010theoretical}
C.~W. Hsu, F.~Sciortino and F.~W. Starr, \emph{Physical review letters}, 2010,
  \textbf{105}, 055502\relax
\mciteBstWouldAddEndPuncttrue
\mciteSetBstMidEndSepPunct{\mcitedefaultmidpunct}
{\mcitedefaultendpunct}{\mcitedefaultseppunct}\relax
\EndOfBibitem
\bibitem[Sciortino \emph{et~al.}(2007)Sciortino, Bianchi, Douglas, and
  Tartaglia]{sciortino2007self}
F.~Sciortino, E.~Bianchi, J.~F. Douglas and P.~Tartaglia, \emph{The Journal of
  Chemical Physics}, 2007, \textbf{126}, 194903\relax
\mciteBstWouldAddEndPuncttrue
\mciteSetBstMidEndSepPunct{\mcitedefaultmidpunct}
{\mcitedefaultendpunct}{\mcitedefaultseppunct}\relax
\EndOfBibitem
\bibitem[De~Michele \emph{et~al.}(2006)De~Michele, Gabrielli, Tartaglia, and
  Sciortino]{de2006dynamics}
C.~De~Michele, S.~Gabrielli, P.~Tartaglia and F.~Sciortino, \emph{J. Phys.
  Chem. B}, 2006, \textbf{110}, 8064\relax
\mciteBstWouldAddEndPuncttrue
\mciteSetBstMidEndSepPunct{\mcitedefaultmidpunct}
{\mcitedefaultendpunct}{\mcitedefaultseppunct}\relax
\EndOfBibitem
\bibitem[Wilber \emph{et~al.}(2009)Wilber, Doye, and Louis]{wilber2009self}
A.~W. Wilber, J.~P.~K. Doye and A.~A. Louis, \emph{Journal of Chemical
  Physics}, 2009, \textbf{131}, 5101\relax
\mciteBstWouldAddEndPuncttrue
\mciteSetBstMidEndSepPunct{\mcitedefaultmidpunct}
{\mcitedefaultendpunct}{\mcitedefaultseppunct}\relax
\EndOfBibitem
\bibitem[Jankowski and Glotzer(2012)]{Jankowski2012}
E.~Jankowski and S.~C. Glotzer, \emph{Soft Matter}, 2012, \textbf{8},
  2852\relax
\mciteBstWouldAddEndPuncttrue
\mciteSetBstMidEndSepPunct{\mcitedefaultmidpunct}
{\mcitedefaultendpunct}{\mcitedefaultseppunct}\relax
\EndOfBibitem
\end{mcitethebibliography}
\end{document}